\documentstyle[11pt,aspconf,epsf]{article}
\def\spose#1{\hbox to 0pt{#1\hss}}
\def\simlt{\mathrel{\spose{\lower 3pt\hbox{$\mathchar"218$}}
     \raise 2.0pt\hbox{$\mathchar"13C$}}}
\def\simgt{\mathrel{\spose{\lower 3pt\hbox{$\mathchar"218$}}
     \raise 2.0pt\hbox{$\mathchar"13E$}}}

\begin{document}

\title{Limits on the star formation rates of $z>2$ damped Ly$\alpha$
systems from H$\alpha$ spectroscopy}
\author{Andrew~J.~Bunker\altaffilmark{1}}
\affil{Department of Astrophysics, University of Oxford,
OX1~3RH, UK}
\author{Stephen~J.~Warren}
\affil{Imperial College, Department of Physics, Blackett Laboratory,
Prince Consort Road, London SW7~2BZ, UK}
\author{D.~L.~Clements}
\affil{Institut
d'Astrophysique Spatiale, B\^{a}timent 121, Universit\'{e} Paris XI,
F-91405 Orsay CEDEX, France}
\author{Gerard~M.~Williger}
\affil{NASA
Goddard Space Flight Center, Code 681, Greenbelt, Maryland 20771, USA}
\author{Paul~C.~Hewett}
\affil{Institute of Astronomy, Madingley Road, Cambridge CB3~0HA, UK}

\altaffiltext{1}{{\em Present address:} NICMOS Postdoctoral Researcher,
Department of Astronomy, University of California at Berkeley, 601
Campbell Hall, Berkeley CA~94720, USA\\ {\tt email:
bunker@bigz.Berkeley.EDU}} 

\begin{abstract}
We present the results of a long-slit $K$-band spectroscopic search
with CGS4 on UKIRT for H$\alpha$ emission from the objects responsible
for high-redshift ($z>2$) damped Ly$\alpha$ absorption systems.  The
objective was to measure the star-formation rates in these
systems. However, no H$\alpha$ emission was detected above our
$3\,\sigma$ limits of $f\simlt10^{-19}$\,W\,m$^{-2}$, corresponding to
star formation rates $\simlt10\,h^{-2}\,{\rm M_{\odot}}\,{\rm yr}^{-1}$
($q_{0}=0.5$).  These upper limits are more meaningful than those from
searches for Ly$\alpha$ emission because the H$\alpha$ line is
unaffected by resonant scattering.  For $q_{0}=0.5$ our limits are in
conflict with the star formation rates predicted under the assumption
that the high-$z$ DLAs are the fully-formed galactic-disk counterparts
of today's massive spiral galaxies. Deeper spectroscopy is needed to
test this picture for $q_{0}=0.0$. A programme of NICMOS imaging
observations currently underway, combined with VLT spectroscopy, will
provide a detailed picture of the link between DLAs and young galaxies.

\end{abstract}


\section{Introduction}

The history of star formation in the Universe is a topic of enormous
current interest ({\em e.g.}, Madau {\em et al.\ }1996). The damped
Ly$\alpha$ absorption systems (DLAs, Wolfe {\em et al.\ }1986) contain
most of the neutral gas in the Universe, and from their redshift
distribution, and the measured column densities, the evolution in the
co-moving density of neutral gas $\Omega_g$ can be measured ({\em e.g.},
Lanzetta {\em et al.\ }1991). Then the analysis of the variation of
$\Omega_g$ with redshift allows the measurement of the history of star
formation in the Universe (Pei \& Fall 1995) provided the consequences
of dust obscuration are accounted for.

This approach to the history of star formation unfortunately tells us
nothing about how galaxies are assembled.  One school of thought has
DLAs being the (large) progenitors of massive spiral disks ({\em e.g.},
Lanzetta {\em et al.\ }1991; Prochaska \& Wolfe 1997). However, M\o ller
\& Warren (1998) have recently shown that the impact parameters of the
few detected galaxy counterparts of high-redshift DLAs are small (in the
context of this debate) and that the space density of the DLAs at
high-redshift is probably much higher than the space density of spiral
galaxies today.

\section{Searches for Star Formation in Damped Systems}

Here we present the results of a spectroscopic survey for
H$\alpha$\,$\lambda$\,656.3\,nm emission from 8 damped absorption
systems at $2.0<z<2.6$, along the line-of-sight to 6 high-redshift
quasars. At these redshifts H$\alpha$ appears in the near-infrared
$K$-window. The results are relevant to the debate on the nature of the
DLAs, for if the DLAs are the counterparts of today's spiral galaxies
the associated H$\alpha$ emission should be detectable. The rate of
depletion of the cosmic density of neutral gas can be used to compute a
universal star formation rate. The average star formation rate in each
DLA depends then on their space density, so that high measured rates of
star formation would provide support for the view that the DLAs are
massive galaxies already in place at high-$z$. A low measured star
formation rate on the other hand would be in agreement with the
hierarchical picture.

There have been extensive searches for Ly$\alpha\,\lambda$\,121.6\,nm
emission from DLAs but with limited success ({\em e.g.}, Smith {\em et
al.\ }1989; Hunstead, Pettini \& Fletcher 1990; Lowenthal {\em et al.\
}1995). This is generally thought to be due to the fact that resonant
scattering greatly extends the path length of Ly$\alpha$ photons
escaping through a cloud of neutral hydrogen so that even very small
quantities of dust can extinguish the line (Charlot \& Fall
1991). Because the effective extinction can be extremely large this has
the consequence that non-detections do not provide any useful
information on the star formation rates in the DLAs. The H$\alpha$ line,
although intrinsically weaker by a factor $\sim 10$, lies at a longer
wavelength where the extinction is smaller, and is not resonantly
scattered. In consequence a search for H$\alpha$ emission from DLAs may
be more efficient than a search for Ly$\alpha$.

\section{Our Near-IR Spectroscopic Survey}

\begin{figure}
\plotone{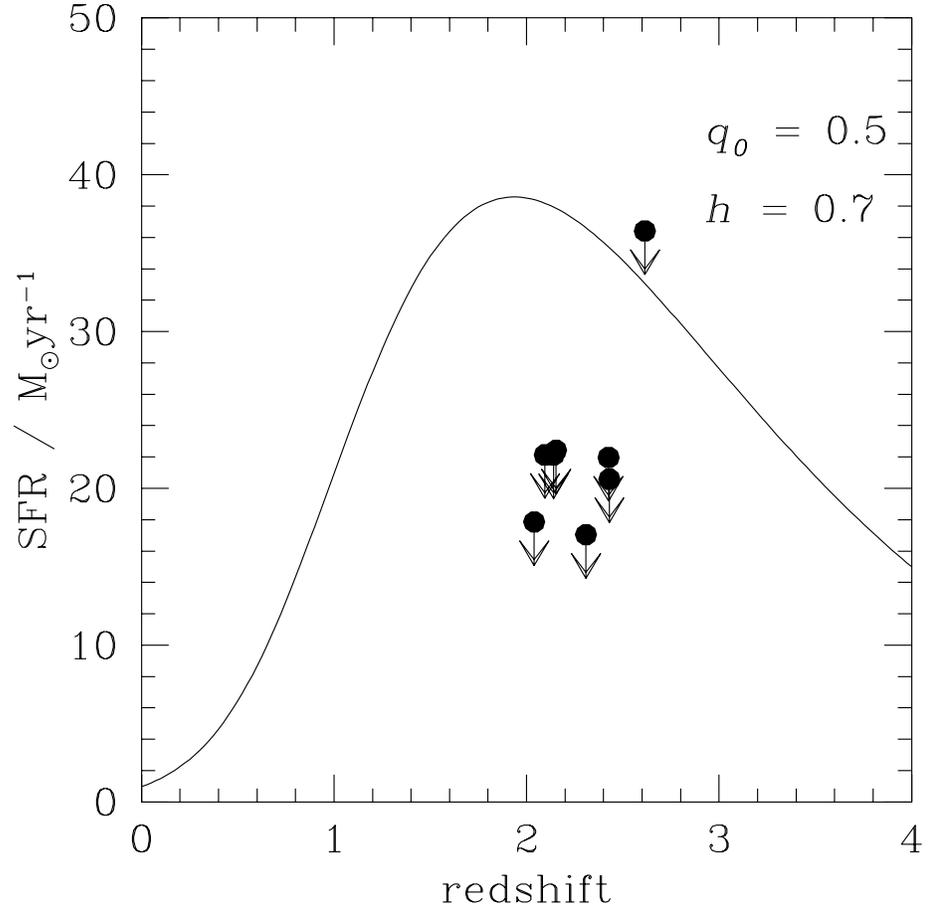}
\caption{The measured $3\,\sigma$ upper limits to the SFRs in ${\rm
M_{\odot}\,yr}^{-1}$ for the sample of 8 DLAs for $q_{\circ}=0.5$. The
curves are the predicted survey-averaged SFRs from the closed-box models
of Pei \& Fall (1995), under the assumption that the DLAs are the
progenitors of present-day spiral galaxy disks. The predictions of the
hierarchical picture will lie below these curves in proportion to the
ratio of the space density of galaxy sub-units at any redshift to the
space density of spiral galaxies today.}
\end{figure}

With the CGS4 spectrograph on the 3.8-m UK Infrared Telescope (UKIRT) we
have undertaken a search for line emission from 8 high-redshift DLAs
($2.0<z<2.6$) near the expected wavelength of H$\alpha$. The
observations and data reduction are detailed in Bunker (1996).  No lines
were detected at $>3\,\sigma$ significance above the quasar
continuum. Our long-slit $K$-band spectra were typically 1-hour each and
used the 2.5-arcsec wide slit ($10\,h^{-1}$\,kpc at $z\approx 2.3$,
$q_{0}=0.5$ \& $h=H_{0}\,/\,100{\rm km\,s}^{-1}\,{\rm Mpc}^{-1}$
throughout).  With a 3-arcsec extraction width, the $3\,\sigma$
upper-limits lie in the range $(1.0 - 1.6)\times 10^{-19}$\,W\,m$^{-2}$
for spectrally-unresolved line emission. The resolution of CGS4 in this
configuration is $650\,{\rm km\,s}^{-1}$ FWHM.

We use the upper limits on H$\alpha$ line luminosities to constrain the
star formation rates in these systems, based on the prescription of
Kennicutt (1983), where a star formation rate (SFR) of $1\,{\rm
M_{\odot}\,yr}^{-1}$ generates a line luminosity in H$\alpha$ of
$11.2\times 10^{33}$\,W. The limits to the SFRs lie in the range $(8.4 -
18)\,h^{-2}\,{\rm M_{\odot}\,yr}^{-1}$, although the conversion between
H$\alpha$ line luminosity and SFR is somewhat uncertain and depends on
the assumed IMF.

\section{Testing the Large Disk Hypothesis}

We compare the measured upper limits to the star formation rates in our
sample against predictions based on the assumption that DLAs are the
progenitors of today's spiral galaxies. We begin with the analysis by
Pei \& Fall (1995) of the observed rate of decline of the cosmic density
of neutral gas $\Omega_g({\rm obs})$ measured from surveys for DLAs. At
any redshift $\Omega_g({\rm obs})$ will be an underestimate of the true
cosmic density $\Omega_g({\rm true})$ because quasars lying behind DLAs
will suffer extinction, and may therefore drop out of the samples of
bright quasars used to find the DLAs. Pei \& Fall correct for this bias,
accounting in a self-consistent manner for the increasing obscuration as
the gas is consumed and polluted as star formation progresses. In this
way they determine the evolution of $\Omega_g({\rm true})$, and so the
SFR per unit volume.

Although the analysis of Pei \& Fall provides the SFR per unit volume at
any redshift, it gives no information on the SFR in individual
galaxies. For the hypothesis of large disks of constant co-moving space
density, assuming that the SFR in a DLA is proportional to the
present-day galaxy luminosity $L(0)$, we can predict the SFRs in the
population of DLAs at any redshift. Figure 1 plots the results of this
calculation, showing the predicted survey-averaged star formation rate
for a sample of DLAs for $q_{\circ}=0.5$. Seven of the eight DLAs lie
below the curve in this plot. The significance of this result is reduced
by the fact that the solid angle over which we have searched for
H$\alpha$ emission is smaller than the expected solid angle of the large
disks. Despite this we would still have expected to detect 2 or 3
systems with average SFRs twice as large as our upper limits. These
results then provide support for the hierarchical picture. For
$q_{\circ}=0.0$ deeper limits are required to distinguish between the
two pictures.  A more detailed treatment of this survey and its
implications is given in Bunker {\em et al.\ }(1998).

\section {Future Work}

A decisive test can be made with the latest generation of near-IR
instrumentation. Deep $H$-band imaging with HST/NICMOS (Warren {\em et
al.}, GO-7824) will reveal whether the galaxies responsible for damped
absorption are indeed in sub-$L^{*}$ pieces, and the opportunities
afforded by the largest ground-based telescopes such as the VLT should
enable the accurate measurement of star formation rates in these
systems. Combined with spectroscopy of metal lines, in this way we will
build up a picture of the history of assembly, gas depletion, and
chemical evolution in the population of damped absorbers.

\acknowledgments We would like to thank the Organizing Committee of the
``NICMOS \& the VLT'' Sardinia meeting.  We are grateful for the
excellent support we received while observing at UKIRT. We thank Mike
Fall, Palle M\o ller \& Hy Spinrad for useful discussions, and Max
Pettini \& Richard McMahon for details of some of the damped systems
included in our survey. AJB was supported by a PPARC studentship while
at Oxford, and a NICMOS-IDT postdoctoral position at U.C.\ Berkeley.

\end{document}